\documentclass{aastex631} % twocolumn

\newcommand{\objname}{2015 FW$_{412}$}

\begin{document}

\title{Discovery of Dust Emission Activity Emanating from Main-belt Asteroid 2015 FW412}

%% A significant change from earlier AASTEX versions is in the structure for 
%% calling author and affiliations. The change was necessary to implement 
%% auto-indexing of affiliations which prior was a manual process that could 
%% easily be tedious in large author manuscripts.
%%
%% Use \affiliation for affiliation information. The old \affil is now aliased to \affiliation. AASTeX v6.31 will automatically index these in the header. When a duplicate is found its index will be the same as its previous entry.
%%
%% The new \altaffiliation can be used to indicate some secondary information such as fellowships. This command produces a non-numeric footnote that is set away from the numeric \affiliation footnotes.  NOTE that if an \altaffiliation command is used it must come BEFORE the \affiliation call, right after the \author command, in order to place the footnotes in the proper location.
%%
%% Use \email to set provide email addresses. Each \email will appear on its own line so you can put multiple email address in one \email call. A new \correspondingauthor command is available in V6.31 to identify the corresponding author of the manuscript. It is the author's responsibility to make sure this name is also in the author list.
%%

%\correspondingauthor{August Muench}
%\email{greg.schwarz@aas.org, gus.muench@aas.org}

\correspondingauthor{Colin Orion Chandler}
\email{coc123@uw.edu}

\author[0000-0001-7335-1715]{Colin Orion Chandler}
\affiliation{Dept. of Astronomy \& the DiRAC Institute, University of Washington, 3910 15th Ave NE, Seattle, WA 98195, USA}
\affiliation{LSST Interdisciplinary Network for Collaboration and Computing, 933 N. Cherry Avenue, Tucson, AZ 85721}
\affiliation{Dept. of Astronomy \& Planetary Science, Northern Arizona University, PO Box 6010, Flagstaff, AZ 86011, USA}

\author[0000-0001-9859-0894]{Chadwick A. Trujillo}
\affiliation{Dept. of Astronomy \& Planetary Science, Northern Arizona University, PO Box 6010, Flagstaff, AZ 86011, USA}

\author[0000-0001-5750-4953]{William J. Oldroyd}
\affiliation{Dept. of Astronomy \& Planetary Science, Northern Arizona University, PO Box 6010, Flagstaff, AZ 86011, USA}

\author[0000-0001-8531-038X]{Jay K. Kueny}
\altaffiliation{National Science Foundation Graduate Research Fellow}
\affiliation{University of Arizona Dept. of Astronomy and Steward Observatory, 933 N. Cherry Avenue Rm. N204, Tucson, AZ 85721, USA}
\affiliation{Lowell Observatory, 1400 W Mars Hill Rd, Flagstaff, AZ 86001, USA}
\affiliation{Dept. of Astronomy \& Planetary Science, Northern Arizona University, PO Box 6010, Flagstaff, AZ 86011, USA}
% \affiliation{Wyant College of Optical Sciences, University of Arizona, 1630 E. University Blvd., Tucson, AZ 85721, USA}

\author[0000-0002-6023-7291]{William A. Burris}
\affiliation{Dept. of Physics, San Diego State University, 5500 Campanile Drive, San Diego, CA 92182, USA}
\affiliation{Dept. of Astronomy \& Planetary Science, Northern Arizona University, PO Box 6010, Flagstaff, AZ 86011, USA}

\author[0000-0001-7225-9271]{Henry H. Hsieh}
\affiliation{Planetary Science Institute, 1700 East Fort Lowell Rd., Suite 106, Tucson, AZ 85719, USA}
\affiliation{Institute of Astronomy and Astrophysics, Academia Sinica, P.O.\ Box 23-141, Taipei 10617, Taiwan}

\author[0000-0002-2204-6064]{Michele T. Mazzucato}
\altaffiliation{Active Asteroids Citizen Scientist}
\affiliation{Royal Astronomical Society, Burlington House, Piccadilly, London, W1J 0BQ, UK}

\author[0000-0002-9766-2400]{Milton K. D. Bosch}
% \affiliation{}
\altaffiliation{Active Asteroids Citizen Scientist}

\author{Tiffany Shaw-Diaz}
\altaffiliation{Active Asteroids Citizen Scientist}

%%%%%%%%%%%%%%%%%%%%%%%%%%%%%%%%%%%%%%%
\begin{abstract}
We present the discovery of activity emanating from main-belt asteroid \objname{}, a finding stemming from the Citizen Science project \textit{Active Asteroids}, a NASA Partner program. We identified a pronounced tail originating from \objname{} and oriented in the anti-motion direction in archival Blanco 4-m (Cerro Tololo Inter-American Observatory, Chile) Dark Energy Camera (DECam) images from UT 2015 April 13, %(Program 2015A-0351, PI Sheppard, observers S. Sheppard, C. Trujillo), 
% and UT 2015 April 
18, 19, 21 and 22.
% (Program 2013B-0536, PI Allen, observers L. Allen, D. James)
% .
Activity occurred near perihelion, consistent with the main-belt comets (MBCs), an active asteroid subset known for sublimation-driven activity in the main asteroid belt; thus \objname{} is a candidate MBC. We did not detect activity on UT 2021 December 12 %(PI Trujillo, observer C.\ Trujillo) 
using the Inamori-Magellan Areal Camera and Spectrograph (IMACS) on the 6.5 m Baade telescope, when \objname{} was near aphelion.
\end{abstract}

%%%%%%%%%%%%%%%%%%%%%%%%%%%%%%%%%%%%%%%%%%%%%%%
\keywords{
Asteroid belt (70), 
Asteroids (72), 
Comae (271), 
Comet tails (274)
}

%%%%%%%%%%%%%%%%%%%%%%%%%%%%%%%%%%%%%%%%%%%%
\section{Introduction} \label{sec:intro}

Active asteroids are rare ($\sim$40) bodies that are found in orbits not normally associated with comets, yet they display comet-like activity such as a tail or coma \citep{jewittActiveAsteroids2015}. A small subset ($\sim$15) of the known active asteroids, the main-belt comets (MBCs), are found to orbit entirely within the main asteroid belt and display comet-like activity indicative of volatile sublimation, such as recurrent activity near perihelion \citep{hsiehMainbeltCometsPanSTARRS12015}. These objects inform us about, for example, the solar system volatile distribution and, consequently, the origins of terrestrial water.

%%%%%%%%%%%%%%%%%%%%%%%%%%%%%%%%%%%%%%%%%%%%%%%%%%%%%%%%%
\section{Methods}
\label{sec:methods}
To find more members of this elusive population we created the Citizen Science project \textit{Active Asteroids}\footnote{\url{http://www.activeasteroids.net}} \citep{chandlerSAFARISearchingAsteroids2018,chandlerSixYearsSustained2019,chandlerCometaryActivityDiscovered2020b,chandlerRecurrentActivityActive2021,chandlerMigratoryOutburstingQuasiHilda2022,chandlerChasingTailsActive2022}, a NASA Partner. Volunteers examine images of known minor planets we extract from publicly available Dark Energy Camera (DECam) data and classify these images as either active or inactive. Volunteers classified images of \objname{} as active, and subsequently we carried out an archival investigation and new telescope observations to further study this object.

%%%%%%%%%%%%%%%%%%%%%%%%%%%%%%%%%%%%%%%%%%%%%%%%%%
\clearpage
\section{Results}
\label{sec:results}

\begin{figure}[h]
    \centering
    \begin{tabular}{ccc}
         \includegraphics[width=0.48\linewidth]{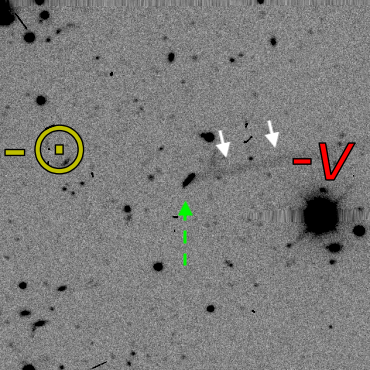} & \includegraphics[width=0.48\linewidth]{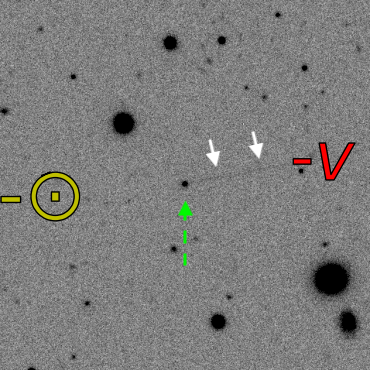} \\
    \end{tabular}
    \caption{\objname{} (green dashed arrow) displays a tail (white arrows) in the anti-motion ($-v$) direction in these images from the DECam on the Blanco 4~m telescope (Cerro Tololo Inter-American Observatory, Chile). Also indicated is the anti-solar (-$\odot$) direction. The FOV is about $126\arcsec \times 126\arcsec$.
    \textbf{Left:} 430~s VR-band, UT 2015 April 13 (Program 2015A-0351, PI Sheppard, observers S. Sheppard, C. Trujillo).
    \textbf{Right:} 40~s VR-band, 2015 April 18 (Program 2013B-0536, PI Allen, observers L. Allen, D. James).
    % \textbf{Right:} 
    }
    \label{fig:activity}
\end{figure}

We identified $\sim$20 archival images of \objname{} (semi-major axis $a=2.76$~au, eccentricity $e=0.16$, inclination $i=13.7^\circ$, perihelion distance $q=2.32$~au, aphelion distance $Q=3.21$~au, Tisserand parameter with respect to Jupiter $T_\mathrm{J}=3.280$; retrieved UT 2023 January 25 from JPL Horizons; \citealt{giorginiJPLOnLineSolar1996}) with clear indications of dust emission activity in the form of a thin tail approximately aligned with the object's orbit plane (Figure \ref{fig:activity}). Notably, \objname{} is a MBC candidate with characteristics consistent with membership of this class, specifically activity occurring near perihelion, and orbiting within the main asteroid belt.% removing per CAT 2/5/2023 COC We shall present additional analysis as part of a forthcoming paper about \textit{Active Asteroids} results.% that presents results from the first year of \textit{Active Asteroids}.
On UT 2015 April 13, when \objname{} was at a heliocentric distance $r_h=2.40$~au (inbound to perihelion), $\alpha=3.1^\circ$ phase angle, and $\nu=320.1^\circ$ true anomaly, the visible angular tail length as projected on the sky was $\sim44\arcsec$, corresponding to $\sim4.5\times10^3$~km at 1.4~au from Earth. We did not detect activity in four 200~s WB4800-7800 filter exposures we acquired with the Inamori-Magellan Areal Camera and Spectrograph on the Magellan 6.5~m Baade telescope on UT 2021 December 12 (PI Trujillo, observer C.\ Trujillo), when \objname{} was at $r_h=3.10$~au, outbound towards aphelion.

%%%%%%%%%%%%%%%%%%%%%%%%%%%%%%%%%%%%%%%%%%%%
\clearpage
\section*{Acknowledgements}
\begin{acknowledgments}
% IMPORTANT: leave a blank line after this one!

\textbf{General:} We thank Dr.\ Mark Jesus Mendoza Magbanua (University of California San Francisco) for his continuous and timely feedback on our project. 

% Note (1/30/2023 COC): all acknowledgement requests for the original discovery were sent out by 2022 June 16. Recently (1/24) I sent out a few more, for classifications that took place subsequently. If anyone responds after the note is published, they will still be included in the actual year 1 publication.
\textbf{Citizen Science:} We thank Elizabeth Baeten (Belgium) for moderating the Active Asteroids forums. We thank our NASA Citizen Scientists that examined \objname{}: 
Al Lamperti	(Royersford, USA), 
Alice Juzumas (São Paulo, Brazil), 
Bill Shaw (Fort William, Scotland), 
Brian K.\ Bernal (Greeley, USA), 
C.\ J.\ A.\ Dukes (Oxford, UK),
Clara Garza (West Covina, USA), 
Dr.\ David Collinson (Mentone, Australia), % @DaveCollinson
Emilio Jose Rabadan Sevilla	(Madrid, Spain), % @emiliozoo 
\texttt{$@$graham\_d} (Hemel Hempstead, UK), 
Graham Mitchell (Chilliwack, Canada), 
Ivan A. Terentev (Petrozavodsk, Russia), 
Jayanta Ghosh (Purulia, India), 
Leah Mulholland	(Peoria IL, USA), 
Martin Welham, (Yatton, UK), % @marwel
Megan Powell (Cobham, UK), % @megpowellx
Michael Jason Pearson, (Hattiesburg MS, USA), 
Michele T.\ Mazzucato (Florence, Italy), 
Sergey Y.\ Tumanov, (Glazov, Russia),
Stikhina Olga Sergeevna (Tyumen, Russia), 
Thorsten Eschweiler (Übach-Palenberg, Germany), 
Tiffany Shaw-Diaz (Dayton, USA), 
and 
Virgilio Gonano	(Udine, Italy) % Wirg78
.
We also thank % super-classifier 
Marvin W.\ Huddleston (Mesquite, USA)% @kc5lei
% and 
% Milton K D Bosch, MD (Napa, USA) % @miltonbosch
. 
Many thanks to Cliff Johnson (Zooniverse) and Marc Kuchner (NASA) for their ongoing guidance.

\textbf{Funding:} This material is based upon work supported by the NSF Graduate Research Fellowship Program under grant No.\ 2018258765 and grant No.\ 2020303693. % removing second sentence 1/31/2023 COC (too long) Any opinions, findings, and conclusions or recommendations expressed in this material are those of the author(s) and do not necessarily reflect the views of the National Science Foundation.  
C.O.C., H.H.H., and C.A.T.\ acknowledge support from the NASA Solar System Observations program (grant 80NSSC19K0869). W.J.O. acknowledges support from NASA grant 80NSSC21K0114. % added per coc 2/5/2023 COC -- note this didn't make into my resubmission this morning, but we still haven't gotten to an editorial phase, so maybe it can still go in.  
This work was supported in part by NSF awards 1950901 (NAU REU program in astronomy and planetary science). % 7/12/2022 COC -- per DET, for WAB?
Computational analyses were run on Northern Arizona University's Monsoon computing cluster, funded by Arizona's Technology and Research Initiative Fund.% This work was made possible in part through the State of Arizona Technology and Research Initiative Program. 

\textbf{Software \& Services:} 
% ``GNU's Not Unix!'' (GNU) Astro \textit{astfits} \citep{Akhlaghi:2015gv} provided command-line FITS file header access.
% C Flexible Image Transport System Input Output (CFITSIO) enabled FITS compression and more \cite{Pence:1999us}. 
World Coordinate System corrections facilitated by \textit{Astrometry.net} \citep{langAstrometryNetBlind2010}. 
This research has made use of 
NASA's Astrophysics Data System, 
%. 
% This research has made use of 
the Institut de M\'ecanique C\'eleste et de Calcul des \'Eph\'em\'erides SkyBoT Virtual Observatory tool \citep{berthierSkyBoTNewVO2006}, 
and 
data and/or services provided by the International Astronomical Union's Minor Planet Center, 
SAOImageDS9, developed by Smithsonian Astrophysical Observatory \citep{joyeNewFeaturesSAOImage2006}. 
% This work made use of the Lowell Observatory Asteroid Orbit Database \textit{astorbDB} \citep{bowellPublicDomainAsteroid1994,moskovitzAstorbDatabaseLowell2021}. 
% COC: already cited in software: This work made use of the \textit{astropy} software package \citep{robitailleAstropyCommunityPython2013}.

\textbf{Facilities \& Instrumentation:} This project used data obtained with the Dark Energy Camera (DECam), which was constructed by the Dark Energy Survey (DES) collaboration. % Cutting off hcere to comply with AAS word limits 1/31/2023 COC
This research uses services or data provided by the Astro Data Archive at NSF's NOIRLab. % Trimming duplicate text of 2nd sentence 1/31/2023 COC: NOIRLab is operated by the Association of Universities for Research in Astronomy (AURA), Inc. under a cooperative agreement with the National Science Foundation. 
Based on observations at Cerro Tololo Inter-American Observatory, NSF’s NOIRLab (NOIRLab Prop. ID 2013B-0536; PI: L. Allen; NOIRLab Prop. ID 2015A-0051; PI: S. Sheppard). % clipping remainder for word count 1/31/2023 COC, which is managed by the Association of Universities for Research in Astronomy under a cooperative agreement with the National Science Foundation.
Magellan observations made use of the Inamori-Magellan Areal Camera and Spectrograph (IMACS) instrument \citep{dresslerIMACSInamoriMagellanAreal2011}; PI C. Trujillo.

\end{acknowledgments}

\vspace{5mm}
\facilities{
% HST(STIS), 
% Swift(XRT and UVOT), 
% AAVSO, 
% CTIO:1.3m, 
% CTIO:1.5m, 
CTIO:4m (DECam), 
Magellan:Baade (IMACS)
% CXO
}

%% Similar to \facility{}, there is the optional \software command to allow authors a place to specify which programs were used during the creation of  the manuscript. Authors should list each code and include either a citation or url to the code inside ()s when available.

\software{
        astropy \citep{robitailleAstropyCommunityPython2013}, 
        % {\tt astrometry.net} \citep{langAstrometryNetBlind2010}, % already in acks 1/31/2023 COC
        % {\tt JPL Horizons} \citep{giorginiJPLOnLineSolar1996}, % already in acks, text 1/31/2023 COC
        {\tt Matplotlib} \citep{hunterMatplotlib2DGraphics2007},
        {\tt NumPy} \citep{harrisArrayProgrammingNumPy2020},
        {\tt pandas} \citep{rebackPandasdevPandasPandas2022}, % limiting to 1 citation, removing older mckinneyDataStructuresStatistical2010 1/31/2023 COC
        {\tt SAOImageDS9} \citep{joyeNewFeaturesSAOImage2006},
        {\tt SciPy} \citep{virtanenSciPyFundamentalAlgorithms2020}
        % {\tt Siril}\footnote{\url{https://siril.org}}, % actually not used in this work (but will be for the full paper) so removing 1/31/2023 COC
        % {\tt SkyBot} \citep{berthierSkyBoTNewVO2006} % cited in acks already
          }

\clearpage
\bibliography{zotero}{}

\begin{thebibliography}{}
\expandafter\ifx\csname natexlab\endcsname\relax\def\natexlab#1{#1}\fi
\providecommand{\url}[1]{\href{#1}{#1}}
\providecommand{\dodoi}[1]{doi:~\href{http://doi.org/#1}{\nolinkurl{#1}}}
\providecommand{\doeprint}[1]{\href{http://ascl.net/#1}{\nolinkurl{http://ascl.net/#1}}}
\providecommand{\doarXiv}[1]{\href{https://arxiv.org/abs/#1}{\nolinkurl{https://arxiv.org/abs/#1}}}

\bibitem[{Berthier {et~al.}(2006)Berthier, Vachier, Thuillot, Fernique,
  Ochsenbein, Genova, Lainey, Arlot, Gabriel, Arviset, Ponz, \&
  Solano}]{berthierSkyBoTNewVO2006}
Berthier, J., Vachier, F., Thuillot, W., {et~al.} 2006, in Astronomical {{Data
  Analysis Software}} and {{Systems XV ASP Conference Series}}, Vol. 351
  ({Orem, UT}: {Astronomical Society of the Pacific}), 367

\bibitem[{Chandler(2022)}]{chandlerChasingTailsActive2022}
Chandler, C.~O. 2022, PhD thesis, Northern Arizona University, {Flagstaff,
  Arizona, USA}

\bibitem[{Chandler {et~al.}(2018)Chandler, Curtis, Mommert, Sheppard, \&
  Trujillo}]{chandlerSAFARISearchingAsteroids2018}
Chandler, C.~O., Curtis, A.~M., Mommert, M., Sheppard, S.~S., \& Trujillo,
  C.~A. 2018, Publications of the Astronomical Society of the Pacific, 130,
  114502, \dodoi{10.1088/1538-3873/aad03d}

\bibitem[{Chandler {et~al.}(2019)Chandler, Kueny, Gustafsson, Trujillo,
  Robinson, \& Trilling}]{chandlerSixYearsSustained2019}
Chandler, C.~O., Kueny, J., Gustafsson, A., {et~al.} 2019, The Astrophysical
  Journal Letters, 877, L12, \dodoi{10/gg3qw6}

\bibitem[{Chandler {et~al.}(2020)Chandler, Kueny, Trujillo, Trilling, \&
  Oldroyd}]{chandlerCometaryActivityDiscovered2020b}
Chandler, C.~O., Kueny, J.~K., Trujillo, C.~A., Trilling, D.~E., \& Oldroyd,
  W.~J. 2020, The Astrophysical Journal Letters, 892, L38, \dodoi{10/gg36xz}

\bibitem[{Chandler {et~al.}(2022)Chandler, Oldroyd, \&
  Trujillo}]{chandlerMigratoryOutburstingQuasiHilda2022}
Chandler, C.~O., Oldroyd, W.~J., \& Trujillo, C.~A. 2022, The Astrophysical
  Journal, 937, L2, \dodoi{10.3847/2041-8213/ac897a}

\bibitem[{Chandler {et~al.}(2021)Chandler, Trujillo, \&
  Hsieh}]{chandlerRecurrentActivityActive2021}
Chandler, C.~O., Trujillo, C.~A., \& Hsieh, H.~H. 2021, The Astrophysical
  Journal, 922, L8, \dodoi{10/gnmckw}

\bibitem[{Dressler {et~al.}(2011)Dressler, Bigelow, Hare, Sutin, Thompson,
  Burley, Epps, Oemler, Bagish, Birk, Clardy, Gunnels, Kelson, Shectman, \&
  Osip}]{dresslerIMACSInamoriMagellanAreal2011}
Dressler, A., Bigelow, B., Hare, T., {et~al.} 2011, Publications of the
  Astronomical Society of the Pacific, 123, 288, \dodoi{10.1086/658908}

\bibitem[{Giorgini {et~al.}(1996)Giorgini, Yeomans, Chamberlin, Chodas,
  Jacobson, Keesey, Lieske, Ostro, Standish, \&
  Wimberly}]{giorginiJPLOnLineSolar1996}
Giorgini, J.~D., Yeomans, D.~K., Chamberlin, A.~B., {et~al.} 1996, American
  Astronomical Society, 28, 25.04

\bibitem[{Harris {et~al.}(2020)Harris, Millman, {van der Walt}, Gommers,
  Virtanen, Cournapeau, Wieser, Taylor, Berg, Smith, Kern, Picus, Hoyer, {van
  Kerkwijk}, Brett, Haldane, {del R{\'i}o}, Wiebe, Peterson,
  {G{\'e}rard-Marchant}, Sheppard, Reddy, Weckesser, Abbasi, Gohlke, \&
  Oliphant}]{harrisArrayProgrammingNumPy2020}
Harris, C.~R., Millman, K.~J., {van der Walt}, S.~J., {et~al.} 2020, Nature,
  585, 357, \dodoi{10.1038/s41586-020-2649-2}

\bibitem[{Hsieh {et~al.}(2015)Hsieh, Denneau, Wainscoat, Sch{\"o}rghofer,
  Bolin, Fitzsimmons, Jedicke, Kleyna, Micheli, Vere{\v s}, Kaiser, Chambers,
  Burgett, Flewelling, Hodapp, Magnier, Morgan, Price, Tonry, \&
  Waters}]{hsiehMainbeltCometsPanSTARRS12015}
Hsieh, H.~H., Denneau, L., Wainscoat, R.~J., {et~al.} 2015, Icarus, 248, 289,
  \dodoi{10.1016/j.icarus.2014.10.031}

\bibitem[{Hunter(2007)}]{hunterMatplotlib2DGraphics2007}
Hunter, J.~D. 2007, Computing in Science \& Engineering, 9, 90,
  \dodoi{10.1109/MCSE.2007.55}

\bibitem[{Jewitt {et~al.}(2015)Jewitt, Hsieh, \&
  Agarwal}]{jewittActiveAsteroids2015}
Jewitt, D., Hsieh, H., \& Agarwal, J. 2015, in Asteroids {{IV}} ({Tucson,
  Arizona}: {University of Arizona Press}), 221--241

\bibitem[{Joye(2006)}]{joyeNewFeaturesSAOImage2006}
Joye, W.~A. 2006, in Astronomical {{Data Analysis Software}} and {{Systems XV
  ASP Conference Series}}, Vol. 351, 574--

\bibitem[{Lang {et~al.}(2010)Lang, Hogg, Mierle, Blanton, \&
  Roweis}]{langAstrometryNetBlind2010}
Lang, D., Hogg, D.~W., Mierle, K., Blanton, M., \& Roweis, S. 2010,
  Astronomical Journal, 139, 1782, \dodoi{10.1088/0004-6256/139/5/1782}

\bibitem[{McKinney(2010)}]{mckinneyDataStructuresStatistical2010}
McKinney, W. 2010, in Python in {{Science Conference}}, {Austin, Texas},
  56--61, \dodoi{10.25080/Majora-92bf1922-00a}

\bibitem[{Reback {et~al.}(2022)Reback, {jbrockmendel}, McKinney, den Bossche,
  Augspurger, Roeschke, Hawkins, Cloud, {gfyoung}, Sinhrks, Hoefler, Klein,
  Petersen, Tratner, She, Ayd, Naveh, Darbyshire, Garcia, Shadrach, Schendel,
  Hayden, Saxton, Gorelli, Li, Zeitlin, Jancauskas, McMaster, W{\"o}rtwein, \&
  Battiston}]{rebackPandasdevPandasPandas2022}
Reback, J., {jbrockmendel}, McKinney, W., {et~al.} 2022, Pandas-Dev/Pandas:
  {{Pandas}} 1.4.2, Zenodo, \dodoi{10.5281/zenodo.6408044}

\bibitem[{Robitaille {et~al.}(2013)Robitaille, Tollerud, Greenfield,
  Droettboom, Bray, Aldcroft, Davis, Ginsburg, {Price-Whelan}, Kerzendorf,
  Conley, Crighton, Barbary, Muna, Ferguson, Grollier, Parikh, Nair,
  G{\"u}nther, Deil, Woillez, Conseil, Kramer, Turner, Singer, Fox, Weaver,
  Zabalza, Edwards, Azalee~Bostroem, Burke, Casey, Crawford, Dencheva, Ely,
  Jenness, Labrie, Lim, Pierfederici, Pontzen, Ptak, Refsdal, Servillat, \&
  Streicher}]{robitailleAstropyCommunityPython2013}
Robitaille, T.~P., Tollerud, E.~J., Greenfield, P., {et~al.} 2013, Astronomy \&
  Astrophysics, 558, A33, \dodoi{10/gfvntd}

\bibitem[{Virtanen {et~al.}(2020)Virtanen, Gommers, Oliphant, Haberland, Reddy,
  Cournapeau, Burovski, Peterson, Weckesser, Bright, {van der Walt}, Brett,
  Wilson, Millman, Mayorov, Nelson, Jones, Kern, Larson, Carey, Polat, Feng,
  Moore, VanderPlas, Laxalde, Perktold, Cimrman, Henriksen, Quintero, Harris,
  Archibald, Ribeiro, Pedregosa, \& {van
  Mulbregt}}]{virtanenSciPyFundamentalAlgorithms2020}
Virtanen, P., Gommers, R., Oliphant, T.~E., {et~al.} 2020, Nature Methods, 17,
  261, \dodoi{10.1038/s41592-019-0686-2}

\end{thebibliography}
\bibliographystyle{aasjournal}

%% This command is needed to show the entire author+affiliation list when the collaboration and author truncation commands are used.  It has to go at the end of the manuscript.
%\allauthors

%% Include this line if you are using the \added, \replaced, \deleted commands to see a summary list of all changes at the end of the article.
%\listofchanges

\end{document}